\documentclass[aps,amsfonts,pre,twocolumn,superscriptaddress,showpacs]{revtex4}
\usepackage[dvips]{graphicx}
\usepackage{amssymb,amsfonts,amsmath}

\newcommand{\uv}{\mathbf{u}}
\newcommand{\xv}{\mathbf{x}}
\newcommand{\dv}{\mathbf{d}}
\newcommand{\ev}{\mathbf{e}}
\newcommand{\fv}{\mathbf{f}}
\newcommand{\tv}{\mathbf{t}}
\newcommand{\Hv}{\mathbf{H}}
\newcommand{\Cv}{\mathbf{C}}
\newcommand{\Dv}{\mathbf{D}}
\newcommand{\lv}{\mathbf{l}}
\newcommand{\bv}{\mathbf{b}}
\newcommand{\sv}{\mathbf{s}}
\newcommand{\qv}{\mathbf{q}}
\newcommand{\gradv}{{\boldsymbol \nabla}}
\newcommand{\remove}[1]{}

\begin{document}
\title{Isostaticity, auxetic response,  surface modes, and
conformal invariance in twisted kagome lattices}
\author{Kai Sun}
\affiliation{Condensed Matter Theory Center and Joint Quantum
Institute, Department of Physics, University of Maryland,
College Park, MD 20742, USA} 
\author{Anton Souslov}
\affiliation{Department
of Physics and Astronomy, University of Pennsylvania,
Philadelphia, PA 19104, USA}
\author{Xiaoming Mao}
\affiliation{Department
of Physics and Astronomy, University of Pennsylvania,
Philadelphia, PA 19104, USA}
\author{T.C. Lubensky}
\affiliation{Department
of Physics and Astronomy, University of Pennsylvania,
Philadelphia, PA 19104, USA}

\begin{abstract}
Model lattices consisting of balls connected by central-force
springs provide much of our understanding of mechanical
response and phonon structure of real materials.  Their
stability depends critically on their coordination number $z$.
$d$-dimensional lattices with $z=2d$ are at the threshold of
mechanical stability and are \emph{isostatic}. Lattices with
$z<2d$ exhibit zero-frequency ``floppy" modes that provide
avenues for lattice collapse.  The physics of systems as
diverse as architectural structures, network glasses, randomly
packed spheres, and biopolymer networks is strongly influenced
by a nearby isostatic lattice. We explore elasticity and
phonons of a special class of two-dimensional isostatic
lattices constructed by distorting the kagome lattice.  We show
that the phonon structure of these lattices, characterized by
vanishing bulk moduli and thus negative Poisson ratios and
auxetic elasticity, depends sensitively on boundary conditions
and on the nature of the kagome distortions. We construct
lattices that under free boundary conditions exhibit surface
floppy modes only or a combination of both surface and bulk
floppy modes; and we show that bulk floppy modes present under
free boundary conditions are also present under periodic
boundary conditions but that surface modes are not. In the the
long-wavelength limit, the elastic theory of all these lattices
is a conformally invariant field theory with holographic
properties, and the surface waves are Rayleigh waves. We
discuss our results in relation to recent work on jammed
systems. Our results highlight the importance of network
architecture in determining floppy-mode structure.
\end{abstract}
\maketitle

\section{Introduction}

Networks of balls and springs or frames of nodes
connected by compressible struts provide realistic models for
physical systems from bridges to condensed solids.  Their
elastic properties depend on their coordination number $z$ --
the average number of nodes each node is connected to.  If $z$
is large enough, the networks are elastic solids whose
long-wavelength mechanical properties are described by a
continuum elastic energy with non-vanishing elastic moduli. If
$z$ is small enough, the networks have deformation modes of
zero energy -- they are floppy.  As $z$ is increased from the
floppy side, a critical value, $z_c$, is reached at which
springs provide just enough constraints that the system has no
zero-energy ``floppy" modes \cite{Thorpe1983} (or mechanisms
\cite{Calladine1978} in the engineering literature), and the
system is \emph{isostatic}. The phenomenon of rigidity
percolation \cite{FengSen1984,JacobsTho1995} whereby a sample
spanning rigid cluster develops upon the addition of springs is
one version of this floppy-to-rigid transition. The
coordination numbers of whole classes of systems, including
engineering structures \cite{Heyman1999,Kassimali2005} (bridges
and buildings), randomly packed spheres near jamming
\cite{LiuNag1998,LiuNag2010a,LiuNag2010b,TorquatoSti2010},
network glasses \cite{Phillips1981,Thorpe1983}, cristobalites
\cite{HammondsWin1996}, zeolites
\cite{Zeolitereview,SartbaevaTho2006}, and biopolymer networks
\cite{WilhelmFre2003,HeussingerFrey2006,HuismanLub2011,BroederszMac2011}
are close enough to $z_c$ that their
elasticity and mode structure is strongly influenced
by those of the isostatic lattice.

\begin{figure}
\centerline{\includegraphics{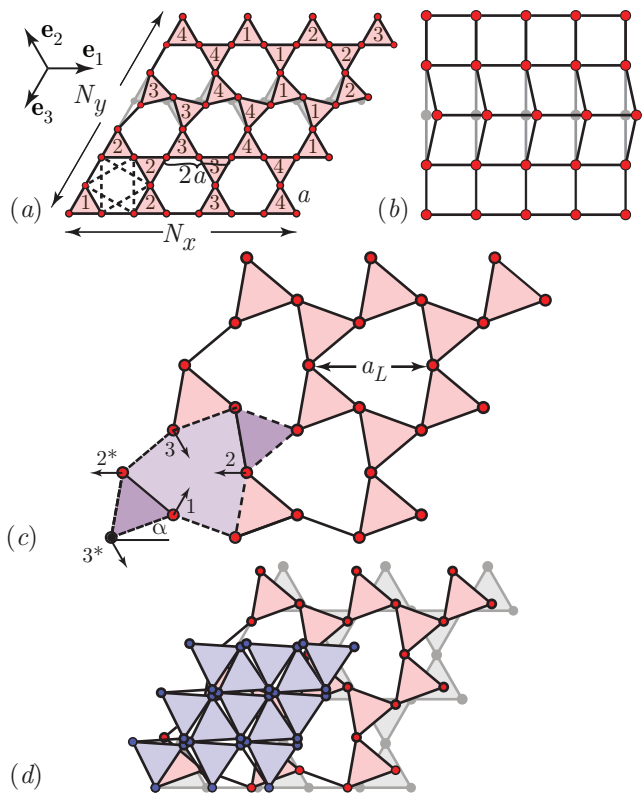}}
\caption{\label{fig:periodic-isostatic} (a) Section of a kagome lattice with
$N_x=N_y=4$ and $N_c=N_x N_y$ three-site unit cells.
Nearest-neighbor bonds, occupied by harmonic springs, are of
length $a$. The rotated row (second row from the top) represents a floppy mode.
Next-nearest neighbor bonds are shown as dotted lines in the
lower left hexagon. The vectors $\ev_1$, $\ev_2$, and $\ev_3$
indicate symmetry directions of the lattice.  The numbers in
the triangles indicate those that twist together under PBCs in
zero modes along the three symmetry direction. Note that
there are only $4$ of these modes. (b) Section of a square
lattice depicting a floppy mode in which all sites along a line
are displaced uniformly. (c) Twisted kagome lattice, with
lattice constant $a_L=2 a \cos \alpha$, derived from the
undistorted lattice by rigidly rotating triangles
through an angle $\alpha$.   A unit cell, bounded by dashed
lines,  is shown in violet. Arrows depict site displacements
for the zone-center, i.e., zero wavenumber, $\phi$ mode which has
zero (nonzero) frequency under free (periodic) boundary
conditions. Sites $1$, $2$, and $3$ undergo no collective
rotation about their center of mass whereas sites $1$, $2^*$,
and $3^*$ do. (d) Superposed snapshots of the twisted lattice
showing decreasing areas with increasing $\alpha$.}
\end{figure}

Though the isostatic point always separates rigid from floppy
behavior, the properties of isostatic lattices are not
universal; rather they depend on lattice architecture. Here we
explore the the unusual properties of a particular class of
periodic isostatic lattices derived from the two-dimensional
kagome lattice by rigidly rotating triangles through an angle
$\alpha$ without changing bond lengths as shown in
Fig.~\ref{fig:periodic-isostatic}. The bulk modulus $B$ of
these lattices is rigorously zero for all $\alpha \neq 0$. As a
result, their Poisson ratio acquires its limit value of $-1$;
when stretched in one direction, they \emph{expand} by an equal
amount in the orthogonal direction: they are maximally auxetic
\cite{Lakes1987, EvansRog1991, LakesChe1991,GreavesRou2011}.
These modes represent collapse pathways
\cite{HutchinsonFle2006,KapkoGue2009} of the kagome lattice.
Modes of isostatic systems are generally very sensitive to
boundary conditions
\cite{Wyartwit2005b,Wyart2005,TorquatoSti2001}, but the degree
of sensitivity depends on the details of lattice structure. For
reasons we will discuss more fully below, modes of the square
lattice, which is isostatic, are in fact insensitive to changes
from free boundary conditions (FBCs) to periodic boundary
conditions (PBCs), whereas those of the undistorted kagome
lattice are only mildly so. The modes of both, however, change
significantly when rigid boundary conditions (RGBs) are
applied. We show here that in all families of the twisted
kagome lattice, modes depend sensitively on whether FBCs, PCBs
or RGBs are applied: finite lattices with free boundaries have
floppy surface modes that are not present in their periodic or
rigid spectrum or in that of finite undistorted kagome
lattices. In the long wavelength limit, the surface floppy
modes, which are present in any $2d$ material with $B=0$,
reduce to surface Rayleigh waves \cite{Landau-elasticity}
described by a conformally invariant energy whose analytic
eigenfunctions are fully determined by boundary conditions.  At
shorter wavelengths, the surface waves become sensitive to
lattice structure and remain confined to within a distance of
the surface that diverges as the undistorted kagome lattice is
approached. In the simplest twisted kagome lattice, all floppy
modes are surface modes, but in more complicated lattices,
including ones with uniaxial symmetry, we construct, there are
both surface and bulk floppy modes.

Arguments due to J.C. Maxwell \cite{Maxwell1864} provide a
criterion for network stability: networks in $d$ dimensions
consisting of $N$ nodes, each connected with central-force
springs to an average of $z$ neighbors, have $N_0=d N -
\frac{1}{2} z N$ zero-energy modes when $z<2d$ (in the absence
of redundant bonds - see below). Of these a number, $N_{\rm
tr}$, which depends on boundary conditions, are trivial rigid
translations and rotations, and the and the remainder are
floppy modes of internal structural rearrangement. Under FBCs
an PBCs, $N_{\rm tr}$ equals $d(d+1)/2$ and $d$, respectively.
With increasing $z$, mechanical stability is reached at the
isostatic point at which $N_0 =N_{\rm tr}$. The Maxwell
argument is a global one; it does not provide information about
the nature of the floppy modes and does not distinguish between
bulk or surface modes.

\section{Kagome zero modes and elasticity}

The kagome lattice of central force springs shown in
Fig.~\ref{fig:periodic-isostatic}(a) is one of many locally
isostatic lattices, including the familiar square lattice
lattice in two dimensions
[Fig.~\ref{fig:periodic-isostatic}(b)] and the cubic and
pyrochlore lattices in three dimensions, with exactly $z=2d$
nearest-neighbor ($NN$) bonds connected to each site not at a
boundary. Under PBCs, there are no boundaries, and every site
has exactly $2d$ neighbors. Finite, $N$-site sections of these
lattices have surface sites with fewer than $2 d$ neighbors and
of order $\sqrt{N}$ zero modes. The free kagome lattice with
$N_x$ and $N_y$ unit cells along its sides
[Fig.~\ref{fig:periodic-isostatic}(a)] has $N=3N_x N_y$ sites,
$N_B =6N_x N_y - 2(N_x + N_y) + 1$ bonds, and $N_0 = 2(N_x +
N_y) - 1$ zero modes, all but three of which are floppy modes.
These modes, depicted in Fig.~\ref{fig:periodic-isostatic}(a),
consist of coordinated counter rotations of pairs of triangles
along the symmetry axes $\ev_1$, $\ev_2$ and $\ev_3$ of the
lattice. There are $N_x$ modes associated with lines parallel
to $\ev_1$, $N_y$ associated with lines parallel to $\ev_3$,
and $N_x+N_y-1$ modes associated with lines parallel to
$\ev_2$.

\begin{figure}
\centerline{\includegraphics{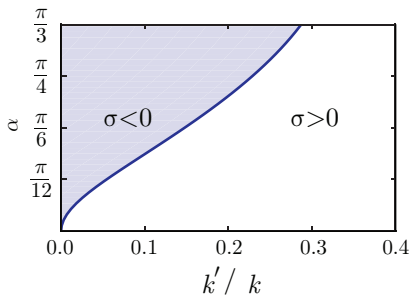}}
\caption{\label{Fig:auxetic-phaseD} phase diagram in
the $\alpha-k'$ plane showing region with negative Poisson ratio $\sigma$.}
\end{figure}

In spite of the large number of floppy modes in the kagome
lattice, its longitudinal and shear Lam\'{e} coefficients,
$\lambda$ and $\mu$, and its Bulk modulus $B = \lambda + \mu$
are nonzero and proportional to the nearest neighbor ($NN$)
spring constant $k$: $\lambda = \mu = \sqrt{3}k/8$, and $B =
\lambda+ \mu = \sqrt{3}k/4$. The zero modes of this lattice can
be used to generate an infinite number of distorted lattices
with unstretched springs and thus zero energy
\cite{KapkoGue2009,SouslovLub2011}. We consider only periodic
lattices, the simplest of which are the twisted kagome lattices
obtained by rotating triangles of the kagome unit cell through
an angle $\alpha$ as shown in
Figs.~\ref{fig:periodic-isostatic}(c) and (d)
\cite{GrimaEva2005,KapkoGue2009}. These lattices have $C_{3v}$
rather than $C_{6v}$ symmetry and, like the undistorted kagome
lattice, three sites per unit cell. As
Fig.~\ref{fig:periodic-isostatic}(d) shows, the lattice
constant of these lattices is $a_L=2 a \cos \alpha$, and their
area $A_{\alpha}$ decreases as $\cos^2 \alpha$ as $\alpha$
increases. The maximum value that $\alpha$ can achieve without
bond crossings is $\pi/3$ so that the maximum relative area
change is $A_{\pi/3}/A_0 = 1/4$. Since all springs maintain
their rest length, there is no energy cost for changing
$\alpha$, and as a result, $B$ is zero for every $\alpha \neq
0$, whereas the shear modulus $\mu= \sqrt{3}k/8$ remains
nonzero and unchanged. Thus, the Poisson ratio $\sigma =
(B-\mu)/(B+ \mu)$ attains its smallest possible value of $-1$.
For any $\alpha \neq 0$, the addition of next-nearest-neighbor
($NNN$) springs, with spring constant $k'$ (or of bending
forces between springs) stabilizes zero-frequency modes and
increases $B$ and $\sigma$. Nevertheless, for sufficiently
small $k'$, $\sigma$ remains negative. Figure
\ref{Fig:auxetic-phaseD} shows the region in the $k'-\alpha$
plane with negative $\sigma$.

\section{Kagome phonon spectrum}

\begin{figure}
\centerline{\includegraphics{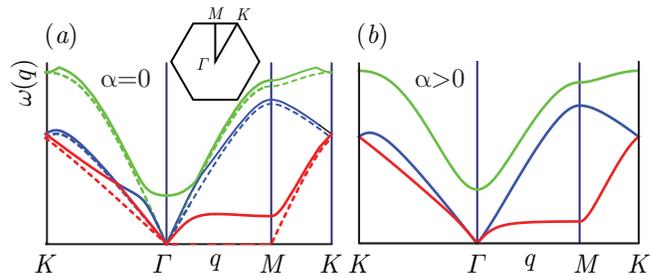}}
\caption{\label{Fig:kagome-spectrum}(a)Phonon spectrum for the undistorted
kagome lattice. Dashed lines depict frequencies at $k'=0$
and full lines at $k'>0$. The inset shows the Brillouin zone
with symmetry points $\Gamma$, $M$, and $K$. Note the line of
zero modes along $\Gamma M$ when $k'=0$, all of which develop
nonzero frequencies for wavenumber $q>0$ when $k'>0$ reaching
$\omega^*\sim \sqrt{k'}$ on a plateau beginning at $q\approx
q^*\sim\sqrt{k'}$ defining a length scale $l^*=1/q^*$. (b) Phonon spectrum
for $\alpha
>0$ and $k'=0$.  The plateau along $\Gamma M$ defines $\omega_{\alpha}
\sim \sqrt{k} |\sin \alpha |$ and its onset at $q_{\alpha}\sim
\omega_{\alpha}$ defines a length $l_{\alpha} \sim 1/|\sin \alpha|$.}
\end{figure}

We now turn to the linearized phonon spectrum of the kagome and
twisted kagome lattices subjected to PBCs.  These conditions
require displacements at opposite ends of the sample to be
identical and thus prohibit distortions of the shape and size
of the unit cell and rotations but not uniform translations,
leaving two rather than three trivial zero modes. The spectrum
\cite{SouslovLub2009} of the three lowest frequency modes along
symmetry directions of the undistorted kagome lattice with and
without $NNN$ springs is shown in
Fig.~\ref{Fig:kagome-spectrum}(a). When $k'=0$, there is a
floppy mode for each wavenumber $\qv \neq 0$ running along the
entire length of the three symmetry-equivalent straight lines
running from $M$ to $\Gamma$ to $M$ in the Brillouin zone [See
inset to Fig.~\ref{Fig:kagome-spectrum}]. When $N_x=N_y$, there
are exactly $N_x-1$ wavenumbers with $\qv\neq 0$ along each of
these lines for a total of $3(N_x -1)$ floppy modes. In
addition, there are three zero modes at $\qv=0$ corresponding
to two rigid translations, and one floppy mode that changes
unit cell area at second but not first order in displacements,
yielding a total of $3N_x$ zero modes rather than the $4N_x-1$
modes expected from the Maxwell count under FBCs. This is our
first indication of the importance of boundary conditions. The
addition of $NNN$ springs endows the floppy modes at $k'=0$
with a characteristic frequency $\omega^* \sim \sqrt{k'}$ and
causes them to hybridize with the acoustic phonon modes
[Fig.~\ref{Fig:kagome-spectrum}(a)] \cite{SouslovLub2009}. The
result is an isotropic phonon spectrum up to wavenumber $q^*
=1/l^* \sim \sqrt{k'}$ and gaps at $\Gamma$ and $M$ of order
$\omega^*$. Remarkably, at nonzero $\alpha$ and $k'=0$, the
mode structure is almost identical to that at $\alpha = 0$ and
$k'>0$ with characteristic frequency
$\omega_\alpha\sim\sqrt{k}|\sin \alpha|$ and length
$l_\alpha\sim 1/\omega_\alpha$. In other words, twisting the
kagome lattice through an angle $\alpha$ has essentially the
same effect on the spectrum as adding $NNN$ springs with spring
constant $|\sin \alpha |^2 k$. Thus under PBCs, the twisted
kagome lattice has no zero modes other than the trivial ones:
it is ``collectively" jammed in the language of references
\cite{TorquatoSti2001,DonevCon2004}, but because it is not
rigid with respect to changing the unit cell size, it is not
strictly jammed.

\section{Mode counting and states of self stress}

\begin{figure}
\centerline{\includegraphics{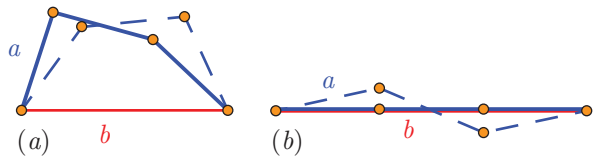}}
\caption{\label{Fig:self-stress}(a)
Ring-network with $b >3a$ showing internal floppy mode. (b)
Ring-network with $b=3a$ showing one of the two
infinitesimal modes.}
\end{figure}

To understand the origin of the differences in the zero-mode
count for different boundary conditions, we turn to an elegant
formulation \cite{Calladine1978} of the Maxwell rule that takes
into account the existence of redundant bonds (i.e., bonds
whose removal does not increase the number of floppy modes
\cite{JacobsTho1995}) and states in which springs can be under
states of self-stress. Consider a ring network in two
dimensions shown in Fig.~\ref{Fig:self-stress} with $N=4$ nodes
and $N_b = 4$ springs with three springs of length $a$ and one
spring of length $b$. The Maxwell count yields $N_0 = 4 = 3+1$
zero modes: two rigid translations, one rigid rotation, and one
internal floppy mode -- all of which are ``finite-amplitude"
modes with zero energy even for finite-amplitude displacements.
When $b=3a$, the Maxwell rule breaks down. In the zero-energy
configuration, the long spring and the three short ones are
colinear, and a prestressed state in which the $b$-spring is
under compression and the three $a$-springs are under tension
(or vice versa) but the total force on each node remains zero
becomes possible. This is called a state of self-stress. The
system still has three finite amplitude zero modes
corresponding to arbitrary rigid translations and rotations,
but the finite-amplitude floppy mode has disappeared. In the
absence of prestress, it is replaced by two ``infinitesimal"
floppy modes of displacements of the two internal nodes
perpendicular of the now linear network.  In the presence of
prestress, these two modes have a frequency proportional to the
square root of the tension in the springs. Thus, the system now
has one state of self stress and one extra zero mode in the
absence of prestress, implying $N_0= 2N - N_B +S$, where $S$ is
the number of states of self stress.

This simple count is more generally valid as can be shown with
the aid of the equilibrium and compatibility matrices
\cite{Calladine1978}, denoted, respectively, as $\Hv$ and
$\Cv\equiv \Hv^T$. $\Hv$ relates the vector $\tv$ of $N_B$
spring tensions to the vector $\fv$ of $dN$ forces at nodes via
$\Hv \cdot \tv = \fv$, and $\Cv$ relates the the vector $\dv$
of $dN$ node displacements to the vector $\ev$ of $N_B$ spring
stretches via $\Cv \cdot \dv = \ev$. The dynamical matrix
determining the phonon spectrum is $\mathbf{D} = k \Hv \cdot
\Hv^T$. Vectors $\tv_0$ in the null space of $\Hv$, ($\Hv \cdot
\tv_0=0$), describe states of self-stress whereas vectors
$\dv_0$ in the null space of $\Cv$ represent displacements with
no stretch $\ev$, i.e., modes of zero energy. Thus the
nullspace dimensions of $\Hv$ and $\Cv$ are, respectively, $S$
and $N_0$. The rank-nullity theorem of linear algebra
\cite{Birkhoff-Mac1998} states that the rank $r$ of a matrix
plus the dimension of its null space equals its column number.
Since the rank of a matrix and its transpose are equal, the
$\Hv$ and $\Cv$ matrices, respectively, yield the relations
$r+S=N_B$ and $r+N_0 = dN$, implying $N_0 = dN- N_B + S$. Under
PBCs, locally isostatic lattices have $z=2d$ exactly, and the
Maxwell rule yields $N_0=0$: there should be no zero modes at
all.  But we have just seen that both the square and
undistorted kagome lattices under PBCs have of order $\sqrt{N}$
zero modes as calculated from the dynamical matrix, which,
because it is derived from a harmonic theory, does not
distinguish between infinitesimal and finite-amplitude zero
modes. Thus, in order for there to be zero modes, there must be
states of self-stress, in fact one state of self-stress for
each zero mode.

In the square lattice under FBCs, $N=N_x N_y$ and $N_B= 2N_x
N_y - N_x -N_y$, there are no states of self stress, and $N_0 =
N_x + N_y$ zero modes depicted in
Fig.~\ref{fig:periodic-isostatic}(b). Under PBCs, the dimension
of the nullspace of $\Hv$ is $S=N_x + N_y$, and there are also
$N_0 = S=N_x+N_y$ zero modes that are identical to those under
FBCs. We have already seen that there are $N_0 = 2(N_x + N_y)
-1$ zero modes in the free undistorted kagome lattice. Direct
evaluations \cite{HutchinsonFle2006} (See Text S1) of the
dimension of the null spaces of $\Hv$ and $\Cv$ for the
undistorted kagome lattice with PBCs yields $S=N_0 = 3N_x$ when
$N_x=N_y$. The zero modes under PBCs are identical to those
under FBCs except that the $2N_x-1$ modes associated with lines
parallel to $\ev_2$ under FBCs get reduced to $N_x$ modes
because of the identification of apposite sides of the lattice
required by the PBCs as shown in
Fig.~\ref{fig:periodic-isostatic}(a). Thus the modes of both
the square and kagome lattices do not depend strongly on
whether FCBs or PBCs are applied. Under RBC's, however, the
floppy modes of both disappear. The situation for the twisted
kagome lattice is different. There are still $2(N_x+N_y)-1$
zero modes under FBCs, but there are only two states of self
stress under PBCs and thus only $N_0 = S = 2$ zero modes, as a
direct evaluation of the null spaces of $\Hv$ and $\Cv$
verifies (See Appendix for details), in agreement with the results
obtained via direct evaluation of the eigenvalues of the
dynamical matrix \cite{SouslovLub2009,GuestHut2003}. All of the
floppy modes under FBCs have disappeared.

\section{Effective theory and edge modes}

An effective long-wavelength energy $E_{\rm eff}$ for the
low-energy acoustic phonons and nearly floppy distortions
provides insight into the nature of the modes of the twisted
kagome lattice. The variables in this theory are the vector
displacement field $\uv ( \xv)$ of nodes at undistorted
positions $\xv$ and the scalar field $\phi(\xv)$ describing
nearly floppy distortions within a unit cell. The detailed form
of $E_{\rm eff}$ depends on which three lattice sites are
assigned to a unit cell.  Figure
\ref{fig:periodic-isostatic}(c) depicts the lattice distortion
$\phi$ for the nearly floppy mode at $\Gamma$ (with energy
proportional to $|\sin \alpha |^2$) along with a particular
representation of a unit cell, consisting of a central
asymmetric hexagon and two equilateral triangles, with $8$
sites on its boundary. If sites $1$, $2$, and $3$ are assigned
to the unit cell, then the distortion $\phi$ involves no
rotations of these sites relative to their center of mass, and
the harmonic limit of $E_{\rm eff}$ depends only on the
symmetrized and linearized strain $u_{ij} = (\partial_i u_j
+\partial_j u_i)/2$ and on $\phi$:
\begin{eqnarray}
E & = & \frac{1}{2}\int d^2 x \Bigl[2 \mu {\tilde u}_{ij}^2 \nonumber\\
& & +  K (\phi + \xi u_{ii})^2 + V (\partial_i \phi)^2 - W \Gamma^{ijk}
u_{ij} \partial_k \phi \Bigr] ,
\label{Eq:full-elastic}
\end{eqnarray}
where ${\tilde u}_{ij} = u_{ij} - \frac{1}{2} \delta_{ij}
u_{kk}$ is the symmetric-traceless stain tensor, $\mu =
\sqrt{3} k/8$, $K = 3 \sqrt{3} \tan^2 \alpha /a^2$, $\xi=a \csc
\alpha /(2 \sqrt{3})$, $W= \sqrt{3}k/4+ O(\alpha^2)$,  and
$V=\sqrt{3}k/8 + O(\alpha^2)$. The last term in which
$\Gamma^{ijk}$ is a third-rank tensor, whose only non-vanishing
components are $\Gamma^{xxx}=-\Gamma^{xyy} = - \Gamma^{yyx} = -
\Gamma^{xyx}=1$, is invariant under operations of the group
$C_{6v}$ but not under arbitrary rotations. The $K \xi \phi
u_{ii}$ term is the only one that reflects the $C_{3v}$ (rather
than $C_{6v}$) symmetry of the lattice. There are several
comments to make about this energy. The gauge-like coupling in
which the isotropic strain $u_{ii}$ appears only in the
combination $(\phi+\xi u_{ii})^2$ guarantees that the bulk
modulus vanishes: $\phi$ will simply relax to $-\xi u_{ii}$ to
reduce to zero the energy of any state with nonvanishing
$u_{ii}$. The coefficient $K$ can be calculated directly from
the observation that under $\phi$ alone, the length of every
spring changes by $\delta a = - \sqrt{3} \phi \sin \alpha$, and
this length change is reversed by a homogenous volume change
$u_{ii}= \delta A_{\alpha}/A_{\alpha} = -2 \delta a/a$. In the
$\alpha\to 0$ limit, $K \to 0$, and the energy reduces to that
of an isotropic solid with bulk modulus $B_0 = \lim_{\alpha \to
0} K \xi^2 = \sqrt{3}k/4$ if the $V$ and $W$ terms, which are
higher order in gradients, are ignored. The $W$ term gives rise
to a term, singular in gradients of $\uv$, when $\phi$ is
integrated out that is responsible for the deviations of the
finite-wavenumber elastic energy from isotropy.  At small
$\alpha$, the length scale $l_{\alpha}$ appears in several
places in this energy: in the length $\xi$ and in the ratios
$\sqrt{\mu/K}$, $\sqrt{V/K}$, and $\sqrt{W/K}$. At length
scales much larger than $l_{\alpha}$, the $V$ and $W$ terms can
be ignored, and $\phi$ relaxes to $-\xi u_{ii}$ leaving only
the shear elastic energy of an elastic solid proportional $\mu
\tilde{u}_{ij}^2$. At length scales shorter than $l_{\alpha}$,
$\phi$ deviates from $-\xi u_{ii}$ and contributes
significantly to the form of the energy spectrum. If $1$, $2^*$
and $3^*$ in Fig.~\ref{fig:periodic-isostatic}(d) are assigned
to the unit cell, then $\phi$ involves rotations relative to
the lattice axes, and the energy develops a Cosserat-like form
\cite{Cosserat1909,Kunin1983} that is a function of $\phi - a
(\gradv \times \uv)_z/2$ rather than $\phi$.

The modes of our elastic energy in the long-wavelength limit
($q l_{\alpha} \ll 1$) are simply those of an elastic medium
with $B=0$. In this limit, there are transverse and
longitudinal bulk sound modes with equal sound velocities $c_T=
\sqrt{\mu/\rho} = (a/2)\sqrt{k/m}$ and $c_L =\sqrt{(B+\mu
)/\rho} \to c_T$, where $m$ is the particle mass at each node
and $\rho$ is the mass density. In addition there are Rayleigh
surface waves \cite{Landau-elasticity} in which there is a
single decay length (rather than the two at $B>0$), and
displacements are proportional to $e^{-q_y y} \cos[q_x x]$ with
$q_y = q_x$ for a semi-infinite sample in the right half plane
so that the penetration depth into the interior is $1/q_x$.
These waves have zero frequency in two dimensions when $B=0$,
and they do not appear in the spectrum with PBCs. Thus this
simple continuum limit provides us with an explanation for the
difference between the spectrum of the free and periodic
twisted kagome lattices. Under FBCs, there are zero-frequency
surface modes not present under PBCs.

Further insight into how boundary conditions affect spectrum
follows from the observation that the continuum elastic theory
with $B=0$ depends only on $\tilde{u}_{ij}$. The metric tensor
$g_{ij} (\xv)$ of the distorted lattice is related to the
strain $u_{ij} ( \xv)$ via the simple relation $g_{ij} ( \xv )
= \delta_{ij} + 2 u_{ij} ( \xv)$; and $\tilde{u}_{ij} =
[g_{ij}( \xv) - \frac{1}{2} \delta_{ij} g_{kk} ( \xv )]/2$,
which is zero for $g_{ij} = \delta_{ij}$, is invariant, and
thus remains equal to zero, under conformal transformations
that take the metric tensor from its reference form
$\delta_{ij}$ to $h(\xv )\delta_{ij}$ for any continuous
function $h(\xv)$. The zero modes of the theory thus correspond
simply to conformal transformations, which in two dimensions
are best represented by the complex position and displacement
variables $z=x+iy$ and $w(z) = u_x(z) + i u_y ( z)$.  All
conformal transformations are described by an analytic
displacement field $w(z)$.  Since by Cauchy's theorem, analytic
functions in the interior of a domain are determined entirely
by their values on the domain's boundary (the ``holographic"
property \cite{Susskind1995}), the zero modes of a given sample
are simply those analytic functions that satisfy its boundary
conditions. For example, a disc with fixed edges ($\uv=0$) has
no zero modes because the only analytic function satisfying
this FBC is the trivial one $w(z)=0$; but a disc with free
edges (stress and thus strain equal to zero) has one zero mode
for each of the analytic functions $w(z) = a_n z^n$ for integer
$n \geq 0$. The boundary conditions $\lim_{x\to \infty} \uv
(x,y) = 0$ and $\uv(x,y) = \uv(x+L, y)$ on a semi-infinite
cylinder with axis along $x$ are satisfied by the function
$w(z) = e^{iq_x z}=e^{i q_x x} e^{-q_x y}$ when $q_x =
2n\pi/L$, where $n$ is an integer. This solution is identical
to that for classical Rayleigh waves on the same cylinder. Like
the Rayleigh theory, the conformal theory puts no restriction
on the value of $n$ (or equivalently $q_x$). Both theories
break down, however, at $q_x =q_c \approx \min (l_\alpha^{-1},
a^{-1})$ beyond which the full lattice theory, which yields a
complex value of $q_y=q_y'+i q_y^{\prime\prime}$, is needed.

\begin{figure}[h]
\centerline{\includegraphics{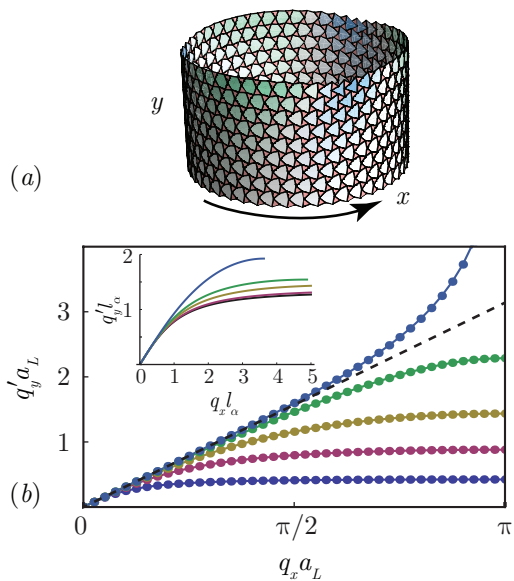}}
\caption{\label{fig:ky-kx}
(a)Lattice distortions for a surface wave on a cylinder, showing
exponential decay of the surface displacements into the bulk. This
figure was constructed by specifying a small sinusoidal modulation
on the bottom boundary and propagating lattice-site positions upward
to the free boundary at the top under the constraint of
of constant lengths and periodic boundary conditions around
the cylinder.  Distortions near and at the top boundary, which have become nonlinear,
are not described by our linearized treatment.
(b) $q_y' a_L$ as a function of $q_x
a_L$ for lattice Rayleigh surface waves for
$\alpha=\pi/20,\pi/10,3\pi/20,\pi/5,\pi/4$, in order from
bottom to top. Smooth curves are the analytic results from a
transfer matrix calculation, and dots are from direct numerical
calculations. The dashed line is the continuum Rayleigh limit
$q_y'=q_x$.  Curves at smaller $\alpha$ break away from this
curve at smaller values of $q_y$ than do those at large
$\alpha$. At $\alpha=\pi/4$, $q_x'$ diverges at $q_y a_L =\pi$.
The inset plots $q_y' l_{\alpha}$ as a function of $q_x
l_{\alpha}$ for different $\alpha$. The lower curve in the
inset(black) is the $\alpha$-independent scaling function of
$q_y l_\alpha$ reached in the $\alpha \to 0$ limit. The other
curves from top to bottom are for $\alpha = \pi/25$, $\pi/12$,
$\pi/9$, and $\pi/6$ (chosen to best present results rather
than to match the curves in the main figure). Curves for
$\alpha < \pi/15$ are essentially indistinguishable from the
scaling limit.  The curve at $\alpha = \pi/6$ stops because
$q_y < \pi/a_L$.}
\end{figure}

Figure \ref{fig:ky-kx}(a) shows an example of a surface wave.
At the bottom of this figure, $u_y(x)$ is an almost perfect
sinusoid.  As $y$ decreases toward the surface, the amplitude
grows, and in this picture reaches the nonlinear regime by the
time the surface at $y=0$ is reached.  Figure
\ref{fig:ky-kx}(b) plots $q_y'$ as a function of $q_x$ obtained
both by direct numerical evaluation and by an analytic transfer
matrix procedure \cite{LeeJoa1981-2} for different values of
$\alpha$ (Text S1). The Rayleigh limit $q_y'=q_x$ is reached
for all $\alpha$ as $q_x \to 0$. Interestingly the Rayleigh
limit remains a good approximation up to values of $q_x$ that
increase with increasing $\alpha$. The inset to
Fig.~\ref{fig:ky-kx}, plots $q_y' l_{\alpha}$ as a function of
$\eta=q_x l_{\alpha}$ and shows that in the limit $\alpha \to
0$ ($l_{\alpha}/a \to \infty$), $q_y'$ obeys an
$\alpha$-independent scaling law of the form $q_y'=
l_{\alpha}^{-1} f( q_x l_{\alpha})$. The full complex $q_y$
obeys a similar equation. This type of behavior is familiar in
critical phenomena where scaling occurs when correlation
lengths become much larger than microscopic lengths. The
function $f(\eta)$ approaches $\eta$  as $\eta \to 0$ and
asymptotes to $4/3$ for $\eta \to \infty$. Thus for $q_x
l_{\alpha} \ll 1$, $q_y' = q_x$ and for $q_x l_{\alpha} \gg 1$,
$q_y' = (4/3) l_{\alpha}^{-1}$. As $\alpha$ increases,
$l_{\alpha}/a$ is no longer much larger than one, and
deviations from the scaling law result. The situation for
surfaces along different directions (e.g., along $x=0$ rather
than $y=0$) is more complicated and will be treated in a future
publication \cite{SunLub2011}.

\section{Other lattices and relation to jamming}

\begin{figure}
\centerline{\includegraphics{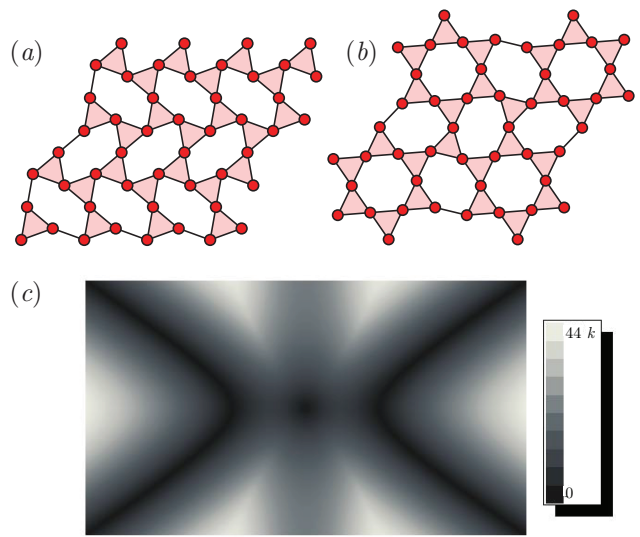}}
\caption{\label{fig:otherauxetic} (a) Kagome-based lattice with pgg space group symmetry and
uniaxial $c_{2v}$ point group symmetry. (b) Lattice with p6
space symmetry but global $C_{6}$ point-group symmetry. (c) Density plot of
the spectrum of the lowest frequency branch of the
pgg uniaxial kagome lattice.  The spectrum is absolutely isotropic near the origin
point $\Gamma$, but it has a zero modes on two symmetry related continuous curves at
large values of wave-number.}
\end{figure}

The $C_{3v}$ twisted kagome lattice is the simplest of many
lattices that can be formed from the kagome and other periodic
isostatic lattices. Figures \ref{fig:otherauxetic} (a) and (b)
show two other examples of isostatic lattices constructed from
the kagome lattice. Most intriguing is the lattice with pgg
symmetry.  Its geometry has uniaxial symmetry, yet its
long-wavelength elastic energy is identical to that of the
$C_{3v}$ twisted kagome lattice, i.e, it is isotropic with a
vanishing bulk modulus, and its mode structure near $\qv=0$ is
isotropic as shown in Fig.~\ref{fig:otherauxetic} (c). Thus,
this system loses long-wavelength zero-frequency bulk modes of
the undistorted kagome lattice to surface modes. However, at
large wavenumber, lattice anisotropy becomes apparent, and
(infinitesimal) floppy bulk modes appear.  Thus in this and
related systems, a fraction of the zero modes of the under FBCs
are bulk modes that are visible under PBCs, and a fraction are
surface modes that are not.

Randomly packed spheres above the jamming transition with
average coordination number $z=2d + \Delta z$ exhibit a
characteristic frequency $\omega^* \sim\Delta z$ and length
$l^* \sim (\Delta z)^{-1}$ and a transition from a Debye-like
($\sim \omega^{d-1}$) to a flat density of states at $\omega
\approx \omega^*$ \cite{OhernNag2003,SilbertNag2005}.  The
square and kagome lattices with randomly added $NNN$ springs
have the same properties \cite{MaoLub2010,MaoLub2011a}. A
general ``cutting" argument \cite{Wyartwit2005b,Wyart2005}
provides a procedure for perturbing away from the isostatic
limit and an explanation for these properties. However, it only
applies provided a finite fraction of the of order $L^{d-1}$
floppy modes of a sample with sides of length $L$ cut from an
isostatic lattice with PBCs are extended, i.e., have wave
functions that extend across the sample rather than remaining
localized either in the interior or at the surface the sample.
Clearly the twisted kagome lattice, whose floppy modes are all
surface modes, violates this criterion; and indeed, the density
of states of the lattice with $\Delta z= 0$ shows
Debye-behavior crossing over to a flat plateau at $\omega
\approx \omega_{\alpha}$. Adding next nearest neighbor bonds
gives rise to a length $l_c \approx (l_{\alpha}^{-1} +
l^{*-1})^{-1}$ and crossover to the plateau at $\omega_c \sim
l_c^{-1}$.  The pgg lattice in Fig.~\ref{fig:otherauxetic}(a),
however, has both extended and surface floppy modes, so its
crossover to the a flat plateau occurs at $\omega \approx
\omega^*$ rather than at $\omega_{\alpha}$ or $\omega_c$.

\section{Connections to other systems}

Our study highlights the rich and remarkable variety of
physical properties that isostatic systems can exhibit. Under
FBCs, floppy modes can adopt a variety of forms, from all being
extended to all being localized near surfaces to a mixture of
the two. Under PBC's, the presence of floppy modes depends on
whether the lattice can or cannot support states of self
stress. When a lattice exhibits a large number of zero-energy
edge modes, its mechanical/dynamical properties become
extremely sensitive to boundary conditions, much as do the
electronic properties of the topological states of matter
studied in
\newpage
\noindent quantum systems
\cite{KaneFisher1997,Jain2007,HasanKane2010,QiZhang2011}. The
zero-energy edge modes observed in our isostatic lattices are
collective modes whose amplitudes decay exponentially from the
edge with a finite decay length, in direct contrast to the very
localized and trivial floppy modes arising from dangling bonds.
We focussed primarily on high-symmetry lattices derived from
the kagome lattice, but the properties they exhibit, namely a
deficit of floppy modes in the bulk and the existence of floppy
surface modes, are shared by any two-dimensional system with a
vanishing bulk modulus (or the equivalent in anisotropic
systems). Three-dimensional analogs of the twisted kagome
lattice \cite{SouslovLub2011} can be constructed by rotating
tetrahedra in pyrochlore and zeolite lattices
\cite{Zeolitereview,SartbaevaTho2006} and in cristobalites
\cite{HammondsWin1996}. These lattices are anisotropic. With
$NN$ forces only, they exhibit a vanishing modulus for
compression applied in particular planes rather than
isotropically, but we expect them to exhibit many of the
properties the two-dimensional lattices exhibit. Finally, we
note that Maxwell's ideas can be applied to spin systems such
as the Heisenberg anti-ferromagnet on the kagome lattice
\cite{MoessnerCha1998,Lawler2011}, and the possibility of
unusual edge states in them is intriguing.

\begin{acknowledgments}
We are grateful for informative conversations with Randall
Kamien, Andea Liu, and S.D. Guest.  This work was supported in
part by NSF under grants DMR- 0804900 (T.C.L. and X.M.), MRSEC
DMR-0520020 (T.C.L. and A.S.) and JQI-NSF-PFC (K. S.).
\end{acknowledgments}

\appendix
\begin{widetext}
\section{Derivation of equilibrium, compatibility, and dynamical
matrices.}

In this section, we will provide some of the details for
calculating the equilibrium and compatibility matrices $\Hv$
and $\Cv$ and the Dynamical matrix $\Dv$ under periodic
boundary conditions. The equilibrium matrix $\Hv$ relates the
$N_B = dN$ dimensional vector of bond tensions $\tv$ to the
$dN$ dimensional vector $\fv$ of forces at the nodes via $\Hv
\cdot \tv = \fv$. We label unit cells by the their position
vectors $\lv=l_1 \ev_1+l_2 \ev_2$, where $l_1$ and $l_2$ are
integers and
\begin{equation}
\ev_n=(\cos \phi_n, \sin \phi_n); \qquad \phi_n = \frac{2\pi (n-1)}{3}
\end{equation}
are the primitive lattice vectors. (Here we take the lattice spacing to
be one, i.e., $2 a \cos \alpha \to 1$.) Each unit cell has three sites,
labeled $1$, $2$, and $3$ as shown in Fig.~\ref{fig:lattice1}, so we
label the $3N_c$ sites by $(\lv, a)$, where $a = 1,2,3$ and $N_c = N/3$
is the number of unit cells. Forces at the nodes are denoted by
$\fv_{\lv,a}$ and their $6N_c = 2N$ components as $\fv_{\lv,a,i}$,
where $i = x,y$. They have to be balanced by stretching forces in the
springs located on the bonds. There are six bonds per cell, which we
can take to be the six bonds in the distorted hexagon shown in
Fig.~\ref{fig:lattice1}, oriented parallel to the unit vectors
\begin{equation}
\bv_n(\alpha ) = (\cos \psi_n,
\sin \psi_n) ; \qquad \psi_n = \frac{(n-1) \pi}{3} + (-1)^n \alpha
\end{equation}
for $n=1,...,6$, where $\alpha$ is the twist angle of the twisted
kagome lattice. When $\alpha = 0$, these vector reduce to the edge
vectors of a symmetric hexagon with $\bv_p(0) = - \bv_{p+3} (0)$.

The $6N_c = 2N$ bonds, labeled with $\lv, n$, are occupied by springs
under tension $t_{\lv, n}$ that exert pulling (for $t_{\lv, n}>0$)
forces, direct along the bond vectors $\pm \bv_n(\alpha)$, on nodes. We
define vectors $\tv_{\lv,n} = t_{\lv, n} \bv_n (\alpha)$. Because we
have periodic boundary conditions, we can express both $\fv_{\lv, a}$
and $\tv_{\lv,n}$ in terms of their Fourier components,
\begin{eqnarray}
\fv_{\lv,a} & = & \frac{1}{\sqrt{N_c}}\sum_\qv e^{i \qv_\cdot \lv} \fv_{\qv,n} \\
\tv_{\lv,n} & = & \frac{1}{\sqrt{N_c}}\sum_\qv e^{i \qv_\cdot \lv} \tv_{\qv,n} .
\end{eqnarray}
The equilibrium matrix is diagonal in $\qv$, so it breaks up into
independent $6\times 6$ blocks for each of the $N_c=N_x N_y$
independent vectors $\qv$.  For simplicity, we consider only the case
$N_x=N_y$ for which $N_c = N_x^2$ and for which the number of sites is
$N=3 N_x^2$.

\begin{figure}
\centerline{\includegraphics{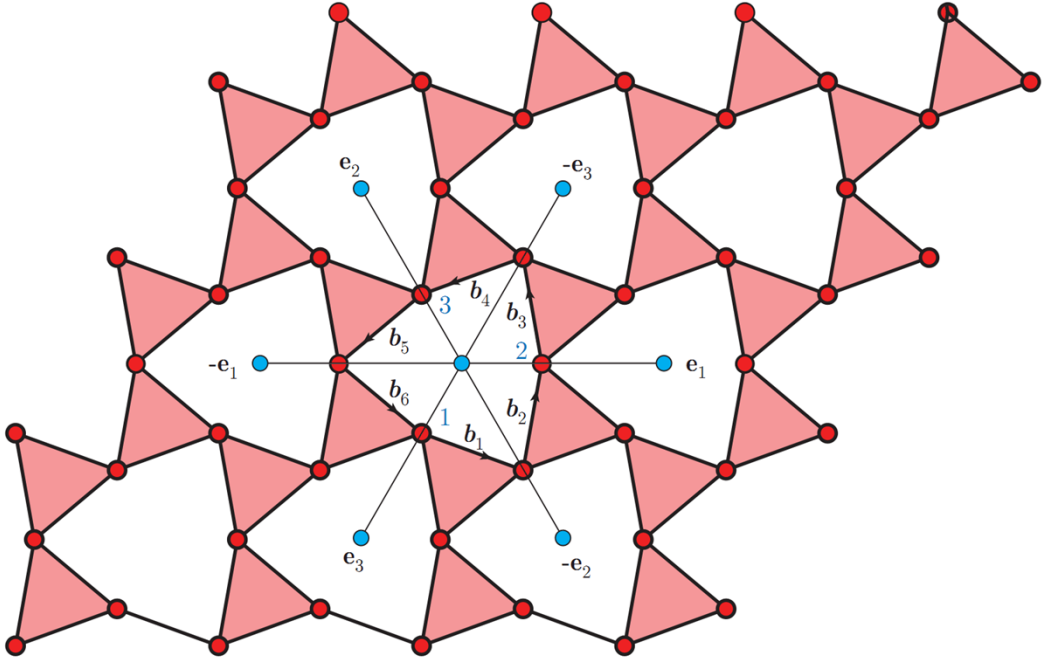}}
\caption{Watch this space}
\label{fig:lattice1}
\end{figure}

\subsection{The Equilibrium Matrix}
There are four bonds incident on each site.  The equations relating
$\fv_{\lv,a}$ to $\tv_{\lv,n}$ follow from Fig.~\ref{fig:lattice1}:
\begin{eqnarray}
\fv_{\lv,1} & = & \tv_{\lv,1} - \tv_{\lv,6} + \tv_{\lv+\ev_3,4} - \tv_{\lv+\ev_3,3} \\
\fv_{\lv,2} & = & \tv_{\lv,3} - \tv_{\lv,2} + \tv_{\lv+\ev_1,6}- \tv_{\lv+\ev_1,5} \\
\fv_{\lv,3} & = & \tv_{\lv,5} - \tv_{\lv,4} + \tv_{\lv+\ev_2,2} - \tv_{\lv+ \ev_2,1} .
\end{eqnarray}
We map $(a,i)$ to a single index $m$ with $(a,x) \to 2a-1$ and
$(a,y)\to 2a$ and then Fourier transform. The result is that
$\Hv$ breaks up into independent $6\times 6$ blocks $\Hv_{nm}
(\qv)$ for each $\qv$. This leads to
\begin{equation}
f_n(\qv) = H_{n,m} (\qv) t_m(\qv) ,
\end{equation}
where
\begin{equation}
\Hv(\qv) =
\begin{pmatrix}
b_{1,x}(\alpha ) & 0 & -e^{i \qv\cdot \ev_3} b_{3,x}(\alpha )& e^{i \qv\cdot \ev_3} b_{4,x}(\alpha )& 0 &-b_{6,x}(\alpha)\\
b_{1,y}(\alpha ) & 0 & -e^{i \qv\cdot \ev_3} b_{3,y}(\alpha ) & e^{i \qv\cdot \ev_3} b_{4,y}(\alpha )& 0 &-b_{6,y}(\alpha)\\
0 & -b_{2,x}(\alpha )& b_{3,x}(\alpha)& 0 &-e^{i \qv\cdot \ev_1} b_{5,x}(\alpha)& e^{i \qv\cdot \ev_1}b_{6,x}(\alpha) \\
0 & -b_{2,y}(\alpha )& b_{3,y} (\alpha)& 0 &-e^{i \qv\cdot \ev_1} b_{5,y}(\alpha)& e^{i \qv\cdot \ev_1}b_{6,y}(\alpha)\\
-e^{i \qv\cdot \ev_2} b_{2,x}(\alpha)& e^{i \qv\cdot \ev_2} b_{1,x}(\alpha)& 0 & b_{5,x}(\alpha) & - b_{4,x}(\alpha)& 0\\
-e^{i \qv\cdot \ev_2} b_{2,y}(\alpha)& e^{i \qv\cdot \ev_2} b_{1,y}(\alpha) & 0 & b_{5,y}(\alpha) & - b_{4,y}(\alpha) & 0
\end{pmatrix} .
\end{equation}
The null space of this matrix is easily calculated with the aid of
Mathematica.

\noindent {\bf Case 1, $\alpha = 0$}:  This is the untwisted
kagome lattice.  When $\qv \neq 0$, the null space of
$\Hv(\qv)$ is empty unless $\qv$ is perpendicular to one of the
primitive lattice vectors:
\begin{equation}
\qv = q_{\perp,n} \ev_{\perp,n} = q( - \sin \phi_n, \cos \phi_n ), \qquad n=1,2,3 ,
\end{equation}
and $\qv\cdot\ev_n = q_{||,n} = 0$, in which case there is a
single vector,
\begin{eqnarray}
t_1(\qv) & = & \sqrt{N_c}\delta_{q_{||,1},0}(1,0,0,e^{-iq_{\perp,1}\, \ev_3 \cdot \ev_{\perp,1}},0,0) \\
t_2(\qv) & = & \sqrt{N_c}\delta_{q_{||,2},0}(0,1,0,0,e^{-iq_{\perp,2}\, \ev_1 \cdot \ev_{\perp,2}},0) \\
t_3(\qv) & = & \sqrt{N_c}\delta_{q_{||,3},0}(0,0,1,0,0,e^{-iq_{\perp,3}\, \ev_2 \cdot \ev_{\perp,3}}) ,
\end{eqnarray}
where $N_c$ is the number of cells, for each $n$ and
$q_{\perp,n}$ in the null space. When $\qv=0$, there are three
vectors, $(1,0,0,1,0,0)$, $(0,1,0,0,1,0)$, and $(0,0,1,0,0,1)$
in the null space.  These are the $q=0$ limits of $t_1(\qv)$,
$t_2(\qv)$, and $t_3(\qv)$, respectively.  We can, therefore,
take the set of null-space vectors of $\Hv$ to be $t_1(\qv)$,
$t_2(\qv)$, and $t_2(\qv)$ for the $N_x$ values of $q$, which
include $q=0$. Linear combinations of these vectors are also
null-space vectors, and we can use them to construct vectors
confined to a single line of bonds. For example the vector
\begin{equation}
t_1 (\lv)  = \frac{1}{N_c} \sum_{q} e^{i\qv\cdot \lv} t_1(\qv)
 = (\delta_{l_y,0},0,0,\delta_{l_y, -\sqrt{3}/2},0,0 )
\end{equation}
corresponds to a state in which springs on all $1$-bond in cells with
centers at $\lv = l_1 \ev_1$ for any $l_1$ and on all $4$-bonds in
cells with centers at $\lv = l_1 \ev_1 + \ev_3 = (l_1-(1/2),
-\sqrt{3}/2)$ are all under the same tension. This is a state of
self-stress in which all bonds along a particular straight line
parallel to the $\ev_1$-axis are under tension. Similar states of
self-stress for any of the $3 N_x$ lines of bonds can be constructed.

\noindent {\bf Case 2, $\alpha >0$}: This is the twisted lattice.  The
null space of $\Hv(\qv)$ is empty for \emph{all} $\qv >0$, and it
contains only two vectors when $\qv = 0$.  These correspond to two
states of self-stress in which the stress the stress on the six bonds
take on both positive and negative values that are identical in all
unit cells.

\subsection{The Compatibility Matrix}
The compatibility matrix $\Cv$ relates the $dN$ displacements
$\dv$ to the $N_b$ stretches $\sv_{\lv,a}$.  The displacements
are labeled the same way as the forces, i.e. as $\dv_{\lv,a}$
and the stretches in the same way as the bond tensions, i.e. as
$s_{\lv,n}$.  The equations relating $\dv_{\lv,a}$ and
$\sv_{\lv,n} \equiv - s_{\lv,n} \bv_n (\alpha) $ are
\begin{eqnarray}
\dv_{\lv-\ev_2,3} - \dv_{\lv,1} & = & \sv_{\lv,1} = - s_{\lv,1} \bv_1 (\alpha) \nonumber \\
\dv_{\lv,2} - \dv_{\lv-\ev_2,3} & = & \sv_{\lv,2} = - s_{\lv,2} \bv_2 (\alpha) \nonumber \\
\dv_{\lv-\ev_3,1} - \dv_{\lv,2} & = & \sv_{\lv,3} = - s_{\lv,3} \bv_3 (\alpha) \nonumber \\
\dv_{\lv,3} - \dv_{\lv-\ev_3,1} & = & \sv_{\lv,4} = - s_{\lv,4} \bv_4 (\alpha) \nonumber \\
\dv_{\lv-\ev_1,2} - \dv_{\lv,3} & = & \sv_{\lv,5} = - s_{\lv,5} \bv_5 (\alpha) \nonumber \\
\dv_{\lv,1} - \dv_{\lv-\ev_1,2} & = & \sv_{\lv,6} = - s_{\lv,6} \bv_6 (\alpha) .
\end{eqnarray}
Fourier transforming and using the same $(n,m)$ notation as for the
equilibrium matrix, we obtain
\begin{equation}
\Cv_{mn}(\qv) d_n(\qv) = s_m(\qv) ,
\end{equation}
where $\Cv_{mn} ( \qv ) = \Hv_{nm} ( - \qv) =
\Hv_{mn}^\dag(\qv)$ as required by the constraint $\Cv =
\Hv^T$.

\subsection{The Dynamical Matrix}
Finally, the dynamical matrix is $\Dv(\qv)= k \Hv(\qv) \cdot
\Cv(\qv)$, whose components are
\begin{eqnarray}
D_{11}(\qv) & = & \frac{1}{2} [\cos (2 \alpha )+4]
\nonumber\\
D_{12}(\qv) & = & = -\frac{1}{2} \sqrt{3} \cos (2 \alpha )
\nonumber\\
D_{13}(\qv) & = & = D_{31}^*(\qv) =- e^{-i \left(\frac{q_x}{2}+\frac{\sqrt{3} q_y}{2}\right)}
\sin ^2\left(\frac{\pi }{6}-\alpha \right)
-e^{-i q_x}\sin ^2\left(\alpha +\frac{\pi}{6}\right)
\nonumber\\
D_{14}(\qv) &=& D_{41}^*(\qv) =\frac{1}{2} e^{-i q_x}\cos \left(\frac{\pi }{6}-2 \alpha \right)
+\frac{1}{2}e^{-i \left(\frac{q_x}{2}+\frac{\sqrt{3} q_y}{2}\right)}
\cos \left(2 \alpha +\frac{\pi }{6}\right)
\nonumber\\
D_{15}(\qv) & = & D_{51}^*(\qv) = -2e^{-\frac{1}{2} i \sqrt{3} q_y} \cos ^2(\alpha )
\cos \left(\frac{q_x}{2}\right)
\nonumber\\
D_{16}(\qv) & = & D_{61}^*(\qv) = i e^{-\frac{1}{2} i \sqrt{3} q_y} \sin (2 \alpha )
\sin \left(\frac{q_x}{2}\right)
\nonumber\\
D_{22}(\qv) & = & 2 \sin ^2(\alpha )+\cos ^2\left(\frac{\pi }{6}-\alpha \right)
+\cos ^2\left(\alpha +\frac{\pi }{6}\right)
\nonumber\\
D_{23}(\qv) & = & D_{32}^*(\qv) =\frac{1}{2}e^{-i q_x} \cos \left(\frac{\pi }{6}-2 \alpha \right)
+\frac{1}{2} e^{-i\left(\frac{q_x}{2}+\frac{\sqrt{3} q_y}{2}\right)}
\cos \left(2 \alpha +\frac{\pi }{6}\right)
\nonumber\\
D_{24}(\qv)&=& D_{42}^*(\qv) =-e^{-i \left(\frac{q_x}{2}+\frac{\sqrt{3} q_y}{2}\right)}
\cos ^2\left(\frac{\pi }{6}-\alpha \right)
-e^{-i q_x}\cos ^2\left(\alpha +\frac{\pi}{6}\right)
\nonumber\\
D_{2,5}(\qv) &=& D_{5,2}^*(\qv) =ie^{-\frac{1}{2} i \sqrt{3} q_y} \sin (2 \alpha )
 \sin \left(\frac{q_x}{2}\right)
\nonumber\\
D_{2,6}(\qv) & = & D_{6,2}^*(\qv) = -2e^{-\frac{1}{2} i \sqrt{3} q_y} \sin ^2(\alpha )
 \cos \left(\frac{q_x}{2}\right)
\nonumber\\
D_{33}(\qv) & = & 2 \sin ^2\left(\frac{\pi }{6}-\alpha \right)+
2 \sin ^2\left(\alpha +\frac{\pi }{6}\right)
\nonumber\\
D_{34}(\qv) &=& D_{43}^* (\qv) = 0
\nonumber\\
D_{3,5}(\qv) & = & D_{5,3}(\qv) =-e^{-i \left(\frac{\sqrt{3} q_y}{2}-\frac{q_x}{2}\right)}
\sin ^2\left(\frac{\pi }{6}-\alpha \right)
-e^{i q_x}\sin ^2\left(\alpha +\frac{\pi}{6}\right)
\nonumber\\
D_{3,6}(\qv) & = & D_{6,3}^*(\qv) = -\frac{1}{2} e^{i q_x}\cos \left(\frac{\pi }{6}-2 \alpha \right)
-\frac{1}{2}e^{i\left(\frac{q_x}{2}-\frac{\sqrt{3} q_y}{2}\right)}
\cos \left(2 \alpha +\frac{\pi }{6}\right)
\nonumber\\
D_{4,4}(\qv) & = & 2 \cos ^2\left(\frac{\pi }{6}-\alpha \right)+
2 \cos ^2\left(\alpha +\frac{\pi }{6}\right)
\nonumber\\
D_{4,5}(\qv) & = & D_{5,4}^*(\qv) = -\frac{1}{2} e^{iq_x}\cos \left(\frac{\pi }{6}-2 \alpha \right)
-\frac{1}{2}e^{i\left(\frac{q_x}{2}-\frac{\sqrt{3} q_y}{2}\right)}
\cos \left(2 \alpha +\frac{\pi }{6}\right)
\nonumber\\
D_{4,6}(\qv) & = & D_{6,4}^*(\qv) = -e^{i \left(\frac{q_x}{2}-\frac{\sqrt{3} q_y}{2}\right)}
\cos ^2\left(\frac{\pi }{6}-\alpha \right)
-e^{i q_x}\cos ^2\left(\alpha +\frac{\pi}{6}\right)
\nonumber\\
D_{5,5}(\qv) & = & \sin ^2\left(\frac{\pi }{6}-\alpha \right)+
\sin ^2\left(\alpha +\frac{\pi }{6}\right)+2 \cos ^2(\alpha )
\nonumber\\
D_{5,6}(\qv) & = & D_{6,5}^*(\qv) = \frac{1}{2}\cos \left(\frac{\pi }{6}-2 \alpha \right)
+\frac{1}{2}\cos \left(2 \alpha +\frac{\pi }{6}\right)
\nonumber\\
D_{6,6}(\qv) & = & 2 \sin ^2(\alpha )+\cos ^2\left(\frac{\pi }{6}-\alpha \right)
+\cos ^2\left(\alpha +\frac{\pi }{6}\right)
\end{eqnarray}
Its eigenvalue spectrum can easily be calculated with the aid
of Mathematica.  The results are shown in Fig.~3 of the main
text.

\section{Surface Modes}
Surface modes decay exponentially into the bulk.  They are
characterized by a wavevector $\qv_{||}$ parallel to the plane
of the surface, their frequency $\omega_s(\qv_{||})$, their
decay length $l(\qv_{||})$ perpendicular to the surface.  In
lattices with nearest-neighbor forces only,
$\omega_s(\qv_{||})$ and $l(\qv_{||})$ can be determined by
setting $\qv=\qv_{||}+\qv_{\perp}$, where $\qv$ is the
wavevector that appears in the dynamical matrix and
$\qv_{\perp}\equiv q_{\perp}\ev_{\perp}$ is the component of
$\qv$ perpendicular to the surface ($\ev_{\perp}$ is the unit
vector perpendicular to the surface), setting $q_{\perp} = i
l^{-1}$, requiring
\begin{equation}
\det [ \omega^2 I - D(\qv_{||}, i l^{-1})] = 0,
\end{equation}
where $I$ is the unit matrix, and that the equation of motion
of the surface layer be satisfied.  In the case of surface
modes of with zero frequency, the relation $\det[D(\qv_{||}, i
l^{-1})]=0$ determines $l$ as a function of $\qv_{||}$. The
surface modes of the twisted kagome lattice have zero
frequency, so we need only solve this equation. We consider
only the case in which the surface is perpendicular to the
$y$-direction and set $q_x = q$ and $r=e^{i\sqrt{3}
q_y/2}\equiv e^{-\sqrt{3}/(2 l)}$.  We find
\begin{equation}
\det[D(q,r)] = \frac{9}{64} a(q)\left[r+r^{-1} - 2g(q)\right]
\left[r+r^{-1} - 2g^*(q)\right] ,
\label{eq:D(q,r)}
\end{equation}
where
\begin{eqnarray}
a(q) & = & 2[2-\cos q +(1-2 \cos q) \cos( 4 \alpha)] \\
g(q) & = & \frac{\cos(q/2)[3 - 2 \cos p + \cos p \cos ( 4 \alpha)]
+ 2 i \sqrt{3} \sin^3 (3 q/2) \sin (4 \alpha)}{2 - \cos q
+(1-2 \cos q) \cos (4 \alpha)}
\end{eqnarray}
Solving for $r$ in Eq.~(\ref{eq:D(q,r)}) and to a lattice
spacing of $2 a \cos \alpha$, we obtain
\begin{equation}
k_y \equiv l^{-1} = \frac{1}{a \sqrt{3} \cos \alpha}
\log[ - g(2 a q \cos \alpha) + \sqrt{g(2 a q \cos \alpha)^2 -1}] .
\end{equation}
This is the function that is plotted in Fig.~5(b) in the text.
Note that $k_y$ has both a real and an imaginary part, but that
in the limit $q\to 0$, $k_y = q$.
\end{widetext}


\end{document}